\documentstyle[prb,aps,epsf,multicol,tabularx]{revtex}

\tighten
\begin{document}


\draft

\title{Friedel phases and phases of transmission amplitudes
in quantum scattering systems}


\author{Tooru Taniguchi and Markus B\"uttiker}

\address{D\'epartement de Physique Th\'eorique, Universit\'e 
de Gen\`eve, CH-1211, Gen\`eve 4, Switzerland}

\date{\today}

\maketitle

\begin{abstract}

We illustrate the relation between the scattering phase appearing in
the Friedel sum rule and the phase of the transmission amplitude for
quantum scatterers connected to two one-dimensional leads.
Transmission zero points cause abrupt phase changes $\pm\pi$ of the
phase of the transmission amplitude.  In contrast the Friedel phase is
a continuous function of energy.  We investigate these scattering
phases for simple scattering problems and illustrate the behavior of
these models by following the path of the transmission amplitude in
the complex plane as a function of energy.  We verify the Friedel sum
rule for these models by direct calculation of the scattering phases
and by direct calculation of the density of states.

\end{abstract}

\vspace{1cm}

\begin{multicols}{2}

\narrowtext

%
%
%
%
%

\newcommand{\scsc}{\scriptscriptstyle}

\section{Introduction}

   The phase is an essential concept in quantum scattering theory.
Some key results and techniques, such as partial wave 
expansion,\cite{book} and the Friedel sum rule\cite{fried} depend in an
explicit manner on the scattering phase.  The Friedel sum rule
connects the density of states to the charge (or a charge difference)
of the system via the phase of the eigenvalues of the scattering
matrix.\cite{fried,lange,langr,dashe}  Since the Friedel phase is
related to the density of states it is also connected to thermodynamic
statistical mechanics quantities\cite{dashe,comme} like the
persistent current.\cite{ackerman}  In addition to the total density
of states the scattering phases also play an important role in the
partial density of states\cite{gaspa} (density of states with a
preselection or postselection of the incident or exiting quantum
channel) and in transport coefficients like capacitances\cite{mello}
and charge relaxation resistances.\cite{mbam}

The principal aim of this work is to investigate the behavior of the
phase in simple scattering problems as they frequently occur in
mesoscopic physics.  In particular we would like to understand which
phases are continuous functions of external parameters (Fermi energy,
magnetic field or Aharonov-Bohm flux) and which phases are permitted
to exhibit jumps as a function of the external parameters.  We
consider coherent quantum scattering systems connected to two
semi-infinite, one-channel leads in the absence of a magnetic field
and without spin-orbit scattering.  Consider for instance the
transmission amplitude $t$ which determines the transmitted current
amplitude if there is an incident current of unit amplitude.  The
transmission amplitude can be expressed in term of its modulus $|t|$
and its phase $\theta^{\scsc (t)}$,

\begin{eqnarray}
        t \equiv \mid t \mid e^{i \theta^{\scsc (t)}}.
\label{phase.t} 
\end{eqnarray}   

\begin{figure}[t]
   \epsfxsize8cm
   \centerline{\epsffile{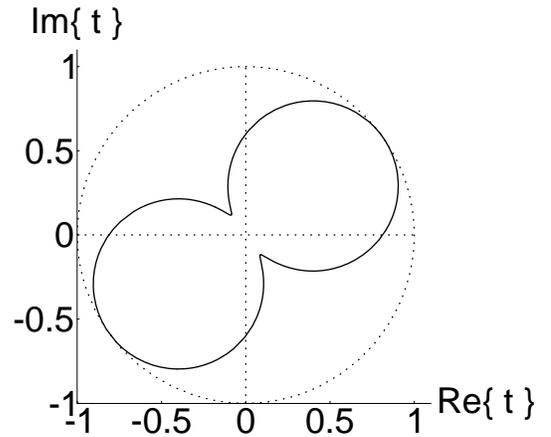}} 
   \vspace{-0.5cm}
   \caption{Transmission amplitude as a function of
   energy for a resonant double-barrier.}
   \label{d.peanu} 
\end{figure}

\noindent If as a function of energy $|t|$ is always positive the path
of the transmission amplitude will encircle the origin of the complex
plane but always stay at a finite distance from it.  An
example of such a behavior is shown in Fig.  \ref{d.peanu}.  As 

\vspace{-1cm}
\vspace{-1cm}
\begin{figure}[t]
   \epsfxsize8cm 
   \centerline{\epsffile{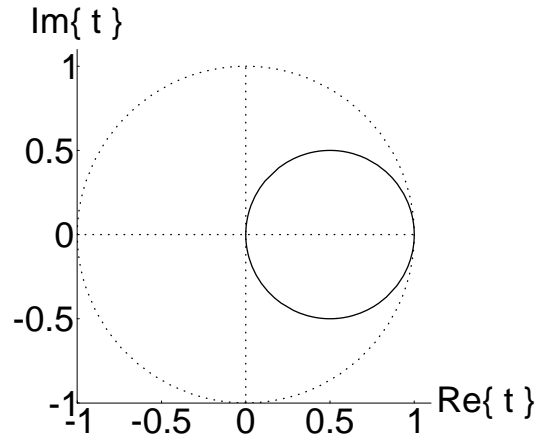}}
   \vspace{-0.5cm}
   \caption{Transmission amplitude as a function 
   of energy for a wire with a side branch.}
   \label{s.egg} 
\end{figure}  

\noindent a consequence the phase $\theta^{\scsc (t)}$ is a continuous function of
the energy.  If on the other hand $|t|$ is zero for a certain value of
the external parameter, the path of the transmission amplitude in the
complex plain will necessarily pass through the origin.  As a
consequence the phase $\theta^{\scsc (t)}$ will jump at this energy by
$\pi$.  Such a behavior is shown in Fig.  \ref{s.egg}.  A specific
model giving rise to the behavior shown in Fig.  \ref{d.peanu} is the
transmission amplitude of a simple one-dimensional resonant tunnel
barrier structure.  The transmission probability never vanishes and
the phase $\theta^{\scsc (t)}$ is a continuous function of energy
which increases in each complete revolution by $2\pi$.  In contrast,
Fig.  2 shows the transmission amplitude for a simple model of a side
branch of finite length attached to a perfect wire.  The transmission
amplitude as function of energy passes through zero and the phase
which increases by $\pi$ through each revolution also exhibits a jump
of $\pi$ (or $-\pi$) bringing the phase back to its origin.

The two models with the very distinctive behavior can be combined and
the evolution of the transmission amplitude of such a combined model
is shown in Fig.  3.  Now the graph shows very many revolutions
through the origin and occasionally a revolution around the origin.
We believe that this reflects the generic behavior of the transmission
amplitude in the complex plane.

Obviously the phase of the transmission amplitude in the second
example, Fig.  2, since it is not a continuous function, can not play
the role of the scattering phase which is used in the Friedel sum
rule.  For a system with a density of states $\rho$ there must exist a
phase which we denote by $\theta^{(f)}$ and which we call the {\em
Friedel phase}, such that its energy derivative is directly related to
the density of states

\begin{equation}
        \frac{\partial \theta^{\scsc (f)}}{\partial E} = \pi \, \rho .
\label{fsum} 
\end{equation}

\noindent Since the density of states should be a continuous function
of the energy for the scattering problems we have in mind, the Friedel
phase must also be a continuous function of energy.  One aim of our
work is to investigate and illustrate the behavior of the different
phases $\theta^{(t)}$ and $\theta^{(f)}$ and to investigate their
connection to the scattering matrix of the problem.

The problem investigated here is of interest in connection with
experiments by Yacoby et al.\cite{yacob} and Schuster et
al.\cite{schus}  In the experiment of Yacoby et al. an Aharonov-Bohm
ring with a quantum dot was investigated in a two terminal geometry.
In a two terminal geometry the Aharonov-Bohm effect exhibits a {\em
parity}:\cite{yeyat}  As function of the Aharonov-Bohm flux the
conductance is either a local minimum at zero flux (positive parity)
or a local maximum (negative parity).  In the experiment of Yacoby et
al.  it was observed that over a sequence of more than a dozen Coulomb
blockade peaks the parity changes at each peak in an identical manner.
It is observed that the parity is positive to the left of the Coulomb
blockade peak and negative to the right of the Coulomb blockade peak.
Such a behavior is incompatible with a simple resonant tunneling model
Fig.  \ref{d.peanu} but would be in accordance with the evolution of
the transmission amplitude shown in Fig.  2.  The second important
experimental fact is that the phase drops by $\pi$ between Coulomb
blockade peaks as shown in the experiment of Schuster et al.  Again
this behavior is compatible with Fig.  2 but not with Fig.
\ref{d.peanu}. 
A number of efforts\cite{yeyat,hacke,brude,yacob.2,hacke.2,deo,oreg,hacke.3,hacke.4,xu,ryu,batin,kang,xu.2,Lee,silve}
have been made to explain the fact that the behavior of the parity is
the same at each peak.  Ref.  \onlinecite{yeyat} suggested a screening
effect, Ref.  \onlinecite{brude} alluded to degeneracies, Refs.
\onlinecite{hacke.2,hacke.3,hacke.4} and \onlinecite{batin} proposed
asymmetric deformation of the dot and repeated tunneling through
exactly the {\em same} state.  The same mechanism due to a dot which
is only semi-chaotic is supported by Silvestrov and Imry\cite{silve}
who point to the large variation in the lifetime of states in a
semi-chaotic dot.  A possible role of zero's of the transmission
probability, Fig.  2, was suggested to us in informal communication by
Levy Yeyati\cite{note1} and has since found interest in a number of
works.\cite{deo,xu,ryu,kang,xu.2,Lee}  We mention here especially the
work by Ryu and Cho\cite{ryu} who investigate an AB ring with a dot
which is also connected to a side branch.  Apart from Ref.
\onlinecite{yeyat} the Friedel sum rule has thus far only found
interest in the recent work of Lee.\cite{Lee}  The work presented
here is in very close connection to the papers by Ryu and Cho and the
paper by Lee.

\begin{figure}[t]
   \epsfxsize8cm  
   \vspace{0cm}
   \centerline{\epsffile{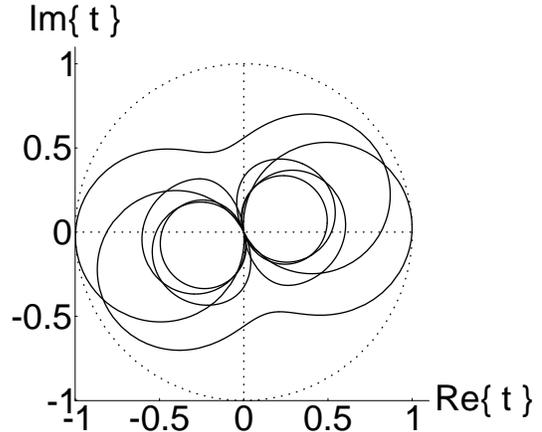}} 
   \vspace{-0.3cm}
   \caption{Transmission amplitude as a function 
   of the particle energy for a model which combines the 
   resonant tunnel barrier and the side branch.}
   \label{c.combin} 
\end{figure}  

It is possible that the experimental observation of identical behavior
over many subsequent conductance peaks is in fact generic (i.e.  is
observed for a fully chaotic quantum dot) and does require a system
with special properties which generates tunneling through the same
state.  The parity of the Ahronov-Bohm effect was observed in a
multi-channel GaAs ring already almost a decade ago in an experiment
by Ford, Washburn and Fowler.\cite{ford}  In this experiment the ring
is coupled to a back gate and the experiments shows wide regions in
the gate voltage - magnetic field plane in which the parity of the
Aharonov-Bohm effect remains the same.

In this work we investigate the various phases of the scattering
matrix and their relation to the phases of the transmission amplitude
and the Friedel phase $\theta^{\scsc (t)}$ comparing in detail two
simple model systems, a one-dimensional resonant tunneling problem and
a perfect wire with a side branch and briefly discuss also the
combination of these two models.


\section{Friedel Phase and Transmission Amplitude Phase}

We consider systems connected to two semi-infinite, one-dimensional
leads.  The system itself can be a multi-dimensional, i.  e.  a
quantum dot.  We consider phase-coherent scattering and neglect
particle-particle interactions and inelastic scattering effects.  In
such systems the scattering matrix $S$ in the particle energy $E$ is
represented as a $2 \times 2$ matrix in which $S_{jj}$ is the
reflection amplitude back into the $j$-th lead for carriers incident
in the $j$-th lead ($j=1,2$) and $S_{jk}$ is the transmission
amplitude from the $k$-th lead to the $j$-th lead ($j \neq k, \,
j=1,2, \, k=1,2$).  The scattering matrix is an unitary matrix,
$S^{-1} = S^{\dagger}$.  This condition guarantees a conservation of
the particle current.  Furthermore it implies that the eigenvalues
of the scattering matrix $S$ are on the unit circle.  Therefore
the eigen value of the scattering matrix $S$ can be represented as
$e^{2 i \xi_j}$ with a real quantity $\xi_j$ $\,$ ($j=1,2$).  Below we
show that we obtain the Friedel sum rule if we define the {\it Friedel
phase} $\theta^{\scsc (f)}$ by 

\begin{eqnarray} 
   \theta^{\scsc (f)}
   \equiv \sum_{j=1}^{2} \xi_j.  
\label{phase.f} 
\end{eqnarray}

\noindent As for any definition of a phase, Eq.  (\ref{phase.f}),
defines the Friedel phase only up to a multiple of $\pi$.  A unitary
$2 \times 2$ matrix can be parameterized as follows,

\begin{eqnarray} \label{Smatr.2}
      S = 
      \left(
      \begin{array}{cc} 
         i e^{i(\theta+\varphi_1)} \sin{\phi} 
         & e^{i(\theta+\varphi_2)} \cos{\phi} 
         \\ e^{i(\theta-\varphi_2)} \cos{\phi} 
         & i e^{i(\theta-\varphi_1)} \sin{\phi}
      \end{array} 
      \right)
\end{eqnarray} 

\noindent with real phases $\theta, \varphi_1, \varphi_2$ and $\phi$.
The eigenvalues of this matrix $\lambda_{j} = \exp({2 i \xi_j})$ are
determined by

\begin{equation}
   \sin(2\xi_{j}-\theta) = \sin{\varphi_{1}} \cdot  \cos{\phi} .
   \label{phase.f.2} 
\end{equation}  

\noindent The eigenvalues are independent of $\varphi_2$.  Furthermore
if $\xi_{-}$ is a solution of Eq.  (\ref{phase.f.2}) then a second
solution is $2\xi_{+} - \theta = \pi - (2\xi_{-} - \theta)$.
Consequently the sum of the two phases of these eigenvalues is
$2\xi_{+} + 2\xi_{-} = \pi + 2 \theta$.  Thus according to Eq.
(\ref{phase.f}) the Friedel phase is given by

\begin{eqnarray}
        \theta^{\scsc (f)} = \theta + \frac{\pi}{2}.
\label{phase.f.3} 
\end{eqnarray} 

\noindent Apart form a constant ${\pi}/{2}$ the Friedel phase is
determined only by the phase $\theta$ of the scattering matrix.  In
particular, as also emphasized by Lee,\cite{Lee} it would be
incorrect to identify the Friedel phase with the argument of the
transmission amplitude.

The derivative of the Friedel phase $\theta^{\scsc (f)}$ with respect
to the particle energy $E$ can be related to the energy derivatives of
the scattering matrix.  The density of states can also be expressed in
terms of the scattering matrix.\cite{dashe}  This gives us a relation
which connects the energy derivatives of the Friedel phase and the
scattering matrix with the density of states,

\begin{eqnarray}\label{fried} 
   \frac{\partial \theta^{\scsc (f)}}{\partial E} 
   && = \frac{1}{4 i} \sum_{j=1}^{2} \sum_{k=1}^{2} \left( 
   S_{jk}^{\ast} \frac{\partial S_{jk}}{\partial E}
   - \frac{\partial S_{jk}^{\ast}}{\partial E} S_{jk} \right) \\ 
   \nonumber
   && = \pi \, \rho
\end{eqnarray}

\noindent Integration of this relation over the energy interval
$[E_{1}, E_{2}]$ gives the Friedel sum rule, which thus states that the
difference in phase $\theta^{\scsc (f)} (E_{2}) -\theta^{\scsc (f)}
(E_{1})$ is equal to the number of particles $N(E_{2},E_{1})$
multiplied by $\pi$ in the system in this energy interval,

\begin{equation}
   \theta^{\scsc (f)} (E_{2}) - \theta^{\scsc (f)} (E_{1}) 
   = \pi N(E_{2},E_{1}).
\end{equation}

\noindent The Friedel phase is thus a {\em a continuous} function of
energy (either $E_{2}$ or $E_{1}$) since the density of states is a
continuous function of energy.

Another phase which is frequently discussed is the argument
$\theta^{\scsc (t)}$ of the transmission amplitude defined by Eq.
(\ref{phase.t}).  Below we will consider only systems with time
reversal invariance.  For such systems the scattering matrix is
symmetric $S_{12} = S_{21}$ and consequently $\varphi_2 = 0$ or
$\varphi_2 = \pi$.  To be definite, we take, $\varphi_2 = 0$.  
In this case the phase of the transmission amplitude is given by

\begin{eqnarray}
       \theta^{\scsc (t)} = \theta 
       + \pi \Theta(\cos{\phi})
\label{phase.t.2} 
\end{eqnarray}  

\noindent where $\Theta(x)$ is the step function of $x$.  In contrast
to the Friedel phase, the phase of the transmission amplitude is thus
in general not a continuous function of energy, but exhibits a jump of
$\pi$ when $\cos{\phi}$ changes sign.  The Friedel phase and the
transmission amplitude phase are not completely independent.  From
Eqs.  (\ref{phase.f.3}) and (\ref{phase.t.2}) we find

\begin{eqnarray}
    \theta^{\scsc (t)} = \theta^{\scsc (f)} - \frac{\pi}{2}
    + \pi \Theta(\cos{\phi}) . 
\label{phase.relat} 
\end{eqnarray} 

\noindent Differentiating Eq.  (\ref{phase.relat}) with respect to
energy (or any other variable of interest) we find,

\begin{eqnarray}
    \Delta\theta^{\scsc (t)} = \Delta\theta^{\scsc (f)} 
    \pm \pi \delta(\cos{\phi}) \Delta\cos{\phi}
\label{phase.relat.2} 
\end{eqnarray}

\noindent where we have used the abbreviation $\Delta = d/dE$.  Eq.
(\ref{phase.relat.2}) shows that only if the condition $\cos{\phi}
\neq 0$ is satisfied for all values of the parameters is the phase of
the transmission amplitudes equal to the Friedel phase.  The condition
$\cos{\phi} \neq 0$ means that the transmission probability $\mid t
\mid^{2}$ is nowhere equal to zero.  Therefore it is the existence of
zero points of the transmission probability which is at the origin of
the difference of the Friedel phase and the phase of the transmission
amplitude.

To be more precise, in addition to a zero point in the transmission
probability, it is also required that the energy derivative of
${\partial\cos{\phi}}/{\partial E}$ is non-zero at the transmission
zero-points.  To see this, we consider phase changes due to the energy
$E$ only, and assume that the number of zero points of the
transmission probability is finite or countable.  The zero points of
the transmission probability determines a sequence of energies which
we denote by $E^{\scsc (n)}$, $n =1, 2, ..$.  Using Eqs.
(\ref{fried}) and (\ref{phase.relat.2}) we obtain

\begin{eqnarray}
   && \frac{\partial \theta^{\scsc (t)}}{\partial E} 
   = \pi \rho + \pi \sum_{n} \mbox{sgn}\left( 
   \left. \frac{\partial 
   \cos{\phi}}{\partial E}\right|_{E=E^{\scsc (n)}}  
   \right) \delta(E-E^{\scsc (n)}) . \nonumber \\ 
   && 
\label{relat.3} 
\end{eqnarray} 

\noindent where the function $\mbox{sgn}(x)$ of $x$ is the sign
function.  It may be noted that the second term of the right-hand side
of Eq.  (\ref{relat.3}) is zero even at a transmission zero point if
${\partial\cos{\phi}}/{\partial E}$ is zero at such a point.  However,
we expect that such cases are unlikely, and we proceed by assuming
that ${\partial\cos{\phi}}/{\partial E}$ is not zero at any
transmission zero point.

To summarize:  The Friedel phase is a continuous function of energy.
On the other hand, the argument of the transmission amplitude, might
exhibit jumps.  These conclusions agree with Lee.\cite{Lee}

\begin{figure}[t] 
   \epsfxsize7.5cm
   \vspace{0.5cm}
   \centerline{\epsffile{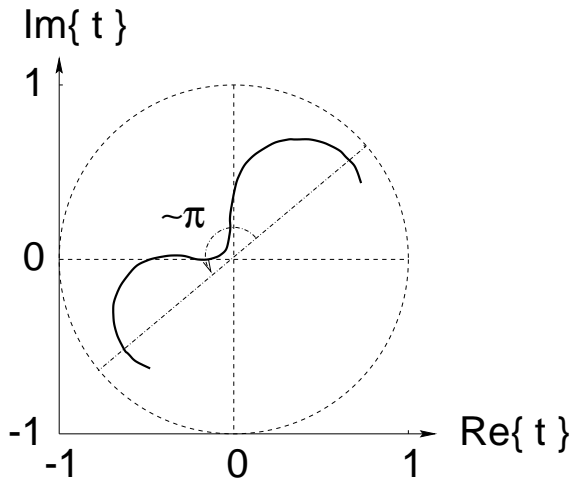}} 
   \vspace{0.7cm}
   \caption{Schematic representation of the path of the transmission 
   amplitude as a function of 
   the particle energy for the case that 
   there is no transmission zero. The energy interval 
   covered is that needed for the transition from 
   one resonant peak (in the transmission probability) 
   to the next resonant peak. The phase changes by about $\pi$.} 
   \label{peanu} 
\end{figure}


\section{Evolution of the Path of the Transmission Amplitude  connecting
consecutive Resonances}  

We now discuss, from a general point of view, the conditions which
lead to the behavior shown in Figs.  \ref{d.peanu}, \ref{s.egg} and
\ref{c.combin}.  We consider a set of parameters for which the
transmission probability shows a series of peaks as function of
energy.  We now ask:  What portion of the path of the transmission
amplitude shown in Fig.  \ref{d.peanu} and Fig.  \ref{s.egg} is traced
out if we increase the energy from its value at a peak in the
transmission probability to a value corresponding to the subsequent
peak in the transmission probability?  Let us denote the energy of the
$n$-th peak in the transmission probability by ${\cal E}_{n}$.  In
addition, let us consider the condition

\begin{eqnarray}
   \int_{{\cal E}_{n}}^{{\cal E}_{n+1}} dE \; \rho \sim 1  
\label{norma} 
\end{eqnarray}

\noindent which according to the Friedel sum rule implies that one
particle is added to the scattering region.  Using this condition and
integrating both sides of Eq.  (\ref{relat.3}) with respect to the
energy from ${\cal E}_{n}$ to ${\cal E}_{n+1}$ we find

\begin{eqnarray}
    && \theta^{\scsc (t)}({\cal E}_{n+1}) 
    -  \theta^{\scsc (t)}({\cal E}_{n}) 
    \sim \left\{
    \begin{array}{ll} 
    \pi & \mbox{case ${\cal A}$}
    \\ 0 \; \mbox{or} \; 2\pi \hspace{0.5cm} 
    & \mbox{case ${\cal B}$} . 
    \end{array} \right.   
\label{dif.phase} 
\end{eqnarray}  
       
\begin{figure}[t] 
   \epsfxsize7.5cm
   \vspace{0.5cm}
   \centerline{\epsffile{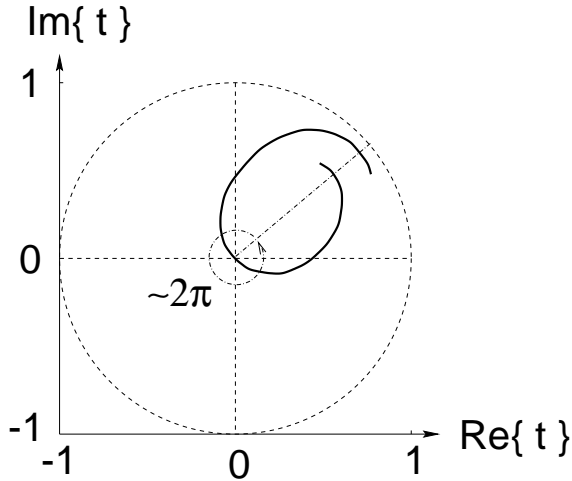}} 
   \vspace{0.7cm}
   \caption{Schematic representation of the path of the 
      the transmission amplitude as a function of 
      the particle energy for the case that there is 
      a transmission zero. The energy interval 
      covered is that needed for the transition from 
      one resonant peak (in the transmission probability) 
      to the next resonant peak.
      The phase changes by about $2\pi$.} 
   \label{egg} 
\end{figure}

\noindent In case ${\cal A}$ there is no transmission zero point in
the energy interval $({\cal E}_{n},{\cal E}_{n+1})$, and in case
${\cal B}$ there is a transmission zero point in the energy interval
$({\cal E}_{n},{\cal E}_{n+1})$.  In case ${\cal A}$ the phase
$\theta^{\scsc (t)}$ evolves by $\pi$ through the consecutive resonant
peaks, so that the resulting path of the transmission amplitude is as
shown in Fig.  \ref{peanu}.  On the other hand, in case ${\cal B}$ the
phase $\theta^{\scsc (t)}$ of the transmission amplitude increases by
$2 \pi$ (or $0$) between the consecutive resonant peaks, so that the
trace of the transmission amplitude is as shown in Fig.  \ref{egg}.

It is clear that Fig.  \ref{d.peanu} is composed of two paths of the
type shown in Fig.  \ref{peanu}, and Fig.  \ref{s.egg} is the result
of a path of the type shown in Fig.  \ref{egg}.  Consequently, there
is a profound difference in the behavior of these two systems:
Whereas, for instance, in a double-barrier scattering problem, we need
to increase the energy over two consecutive states to re-arrive at the
starting point, for the wire with a side branch an energy increase
over one state only is sufficient to bring the system back to the same
point.  We also remark that Fig.  \ref{c.combin} is composed of
combinations of the paths shown in Figs.  \ref{peanu} and \ref{egg}.

Clearly it would be desirable to classify all the possible paths that
are taken in the complex plane as we proceed from one transmission
peak to another.  Here we have emphasized only two paths, namely those
shown in Fig.  \ref{peanu} and \ref{egg}.  These two possibilities are
a direct consequence of the condition Eq.  (\ref{norma}), but of
course there might exist scattering problems which do not obey this
condition.

To summarize this Section:  We have shown that the addition of a
particle to the system can lead to at least two very distinct paths of
the transmission amplitude in the complex plane.  For paths of type
${\cal A}$ (see Fig.  \ref{peanu}) the parity of the Aharonov-Bohm
effect in a two terminal geometry is {\em out of phase} on consecutive
resonant peaks, whereas for a scatterer of type ${\cal B}$ (see Fig.
\ref{egg}) the Aharonov-Bohm oscillations are {\em in phase}.  With
respect to the experiment of Yacoby et al.  it is an in-phase behavior
that is needed to explain the data.

\section{Examples} 

In this section we present the calculations which lead to 
Figs. \ref{d.peanu}, \ref{s.egg} and \ref{c.combin}
for the transmission amplitude. 
In addition we present results for the density of states 
as a function of energy which we compare 
with the transmission probability. 

\subsection{The resonant double-barrier}

\begin{figure}[t]
   \epsfxsize7.7cm
  \vspace{0.5cm}
  \centerline{\epsffile{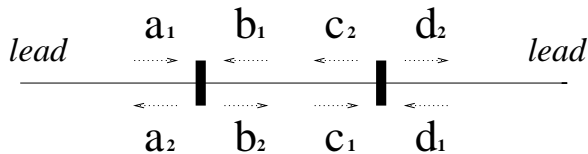}} 
     \vspace{0.5cm}
  \caption{Current amplitudes in the double-barrier model.} 
  \label{doubl} 
\end{figure}

The first example is a one-dimensional double-barrier model, 
which consists of two consecutive potential barriers which 
scatter particles moving along the $x$-axis. 
A schematic illustration is shown in Fig. \ref{doubl}.  
\noindent 
We assume that the two potential barriers in this model are 
identical and that each potential barrier is symmetric with respect 
to particles incident from the left and the right. 
The current amplitudes $a_j$, $b_j$, $c_j$ 
and $d_j$, $j=1,2$ which are shown in Fig. \ref{doubl}, are 
connected by the following relations, 
\begin{eqnarray} 
      \left( 
      \begin{array}{c} 
         a_2 \\ b_2
      \end{array}
      \right)
      = 
      \left(
      \begin{array}{cc} 
         \tilde{r} & \tilde{t} \\ \tilde{t} & \tilde{r} 
      \end{array} 
      \right)
      \left( 
      \begin{array}{c}
         a_1 \\ b_1
      \end{array}
      \right)
   \label{doubl.1}  
\end{eqnarray} 
   \vspace{-0.5cm}
\begin{eqnarray}   
   \left( 
      \begin{array}{c} 
         c_2 \\ d_2
      \end{array}
      \right)
      = 
      \left(
      \begin{array}{cc} 
         \tilde{r} & \tilde{t} \\ \tilde{t} & \tilde{r} 
      \end{array} 
      \right)
      \left( 
      \begin{array}{c}
         c_1 \\ d_1
      \end{array}
      \right) .
   \label{doubl.3} 
\end{eqnarray}

\noindent Here, $\tilde{r}$ and $\tilde{t}$ are the reflection and the
transmission amplitude of the potential barriers, respectively.  The
amplitudes in the well are related by $b_{1} = \tau c_{2}$ and $c_{1}
= \tau b_{2}$, where $\tau = \exp(i\varphi)$ is the transmission
amplitude of the well and $\varphi$ the phase accumulated by a one
time traversal of the well.

We have to find the scattering matrix $S$ 
which relates the outgoing current amplitudes 
$a_2 , d_2$ to the incoming current amplitudes
$a_1, d_1$. A little algebra gives    
   \begin{eqnarray} 
      S = 
      \left(
\begin{array}{cc} 
         \tilde{r} & 0
         \\ 0 & \tilde{r}
      \end{array} 
      \right)  + 
      \frac{\tilde{t}^2 \tau}{1-\tilde{r}^2\tau^2} 
      \left(
      \begin{array}{cc} 
          \tilde{r}\tau & 1 \\ 1 & \tilde{r}\tau 
      \end{array} 
      \right) .  
    \label{d.smatr} 
\end{eqnarray}

\noindent To proceed we parameterize the scattering matrix of the
single barrier also in terms of angular variables (see Eq.
(\ref{Smatr.2})).  We take $\tilde{r} = i e^{i\tilde{\theta}}
\sin{\tilde{\phi}}$ and $\tilde{t} = e^{i\tilde{\theta}}
\cos{\tilde{\phi}}$ with real angles $\tilde{\theta}$ and
$\tilde{\phi}$.  In our model the potential is uniform (except for the
two barriers) and $\varphi$ is thus connected to the energy $E$ of a
scattering particle via $\varphi = k l$ where $l$ is the distance
between the potential barriers and $k = \sqrt{2mE}/\hbar$ is the wave
vector of the particle away from the barriers.  We assume that the
quantities $\tilde{\theta}$ and $\tilde{\phi}$ are independent of the
energy $E$.  This implies in particular that there is no particle
density inside each potential barrier.  With these specifications the
transmission amplitude $t$ is given by

\begin{eqnarray}
   t = \frac{e^{i(kl+ 2\tilde{\theta})}
   \cos^{2}{\tilde{\phi}} }{ 1
   + e^{2i(kl+ \tilde{\theta})} 
   \sin^{2}{\tilde{\phi}}} .
   \label{d.trans.2} 
\end{eqnarray}
\noindent  
The transmission amplitude is a function of energy only through 
the energy dependence of the wave vector $k$.
This result is used 
to give the transmission amplitude in the complex plane as a function 
of energy in Fig. \ref{d.peanu}. The parameters chosen for  
Fig. \ref{d.peanu} are  $\tilde{\theta}=2.2$ and $\tilde{\phi}=2.1$.

\subsection{Phases and density of states in the double-barrier model} 

For comparison with the wire connected to a side branch, we now 
examine the phases in the double barrier scattering problem and 
the density of states. To find the Friedel phase we use 
Eq. (\ref{fried}) and find from  Eqs. (\ref{d.smatr}), 
   
   \begin{eqnarray} 
      \frac{\partial \theta^{\scsc (f)}}{\partial E} = 
      \frac{\partial \varphi}{\partial E}
      \frac{1-\mid\tilde{r}\mid^{4}}{\mid 1 
      -\tilde{r}^{2}\tau^{2} \mid^{2}}.   
   \label{d.derivfried} \end{eqnarray}   
  
\noindent 
The derivative of the phase $\theta^{\scsc (t)}$ of the transmission 
amplitude is found from $\tan \theta^{\scsc (t)}
=\mbox{Im}\{t\}/\mbox{Re}\{t\}$, 

\begin{eqnarray} 
      \frac{\partial \theta^{\scsc (t)}}{\partial E} = 
      \left\{1 + \left( \frac{\mbox{Im}\{t\}}{\mbox{Re}
      \{t\}} \right)^{2} \right\}^{-1} 
      \frac{\partial}{\partial E} 
      \frac{\mbox{Im}\{t\}}{\mbox{Re}\{t\}}.
   \label{d.derivtrans} 
\end{eqnarray}

\noindent 
Substituting Eq. (\ref{d.trans.2}) into 
Eq. (\ref{d.derivtrans}) we can show that for the resonant double barrier
the derivatives of the two phases are identical, 

\begin{eqnarray}
      \frac{\partial \theta^{\scsc (t)}}{\partial E} = 
      \frac{\partial \theta^{\scsc (f)}}{\partial E}. 
   \label{d.phaserelat} 
\end{eqnarray} 

\begin{figure}[t]
\epsfxsize7.7cm
\centerline{\epsffile{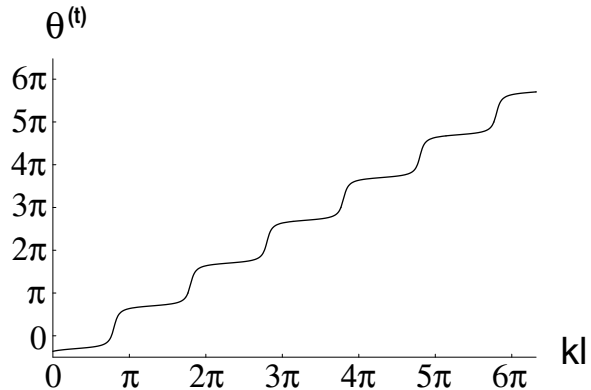}}  
\caption{Phase of the transmission amplitude as a function 
of $kl (= l\sqrt{2mE}/\hbar)$ for the double-barrier model. } 
\label{d.phase} 
\end{figure}

\noindent This result just restates Eq.  (\ref{phase.relat.2}) for the
double-barrier model.  The phase $\theta = \theta^{\scsc (t)} =
\theta^{\scsc (f)} -\pi/2$ for the double barrier model as a function
of $kl$ (wave-vector times well width) is shown in Fig.
\ref{d.phase}.  The specific parameters are $\tilde{\theta}=2.2$ and
$\tilde{\phi}=2.1$.  Thus the phase shows a step like behavior.  It is
nearly constant as a function of $k$ and increases sharply when $k$,
respectively the energy $E$, coincides with a resonant state.  To show
this, we now examine the density of states.

For a perfect one-dimensional wire the density of states per unit
length and in a small interval of wave vectors is $dn/dk = 1/(2\pi)$
for carriers moving to the right.  Since $dE/dk = \hbar v$ where $v$
is the velocity of the carrier, the density of states per unit length
and in a small energy interval is $\nu \equiv dn/dE = 1/(hv)$.  We are
interested in the density of states in the region between the two
barriers.  Denoting the scattering states with unit amplitude of
incident carriers from the left by $\Phi_1(x)$ and the scattering
state of carriers with unit amplitude of carriers incident from the
right by $\Phi_2(x)$, the local density of states can be expressed 
by\cite{MB88}

   \begin{eqnarray}
      \nu (x) = \sum_{\mu=1}^{2} \frac{1}{hv} 
      \mid \Phi_{\mu}(x) \mid^{2} .
   \label{nudensity} 
   \end{eqnarray}
\noindent 
The density of states in the well region (between the two barriers) 
is the integral over the local density of states   
$\rho = \int_{-l/2}^{l/2} dx \nu (x)$. 
With an explicit calculation 
of the scattering states, we obtain 
   \begin{eqnarray}
      \rho =  \frac{2l}{hv} \left| \frac{\tilde{t}}{1-\tilde{r}^{2} 
      \tau^{2}} \right|^{2} \left( 1+\mid\tilde{r}\mid^{2} 
      + 2 \mbox{Re}\{\tilde{r}\tau\}
      \frac{\sin\varphi}{\varphi} \right). 
   \label{d.densi.2} 
   \end{eqnarray}

\noindent Here, for simplicity, we assume that the phase increment of
a barrier traversal is zero.  We have considered here the density of
states which are obtained in terms of energy derivatives of phases or
scattering matrices (as it is widely done).  Such derivatives do not
naturally lead to an answer for a spatially local density of states.
The density of states in the region between the two barriers, is
however, a local question.  A rigorous procedure to obtain local
densities is via derivatives of phases and scattering matrices with
respect to local potentials.\cite{gaspa}  The discussion given here
(in terms of energy derivatives) is correct only up to WKB like
corrections.  The last term in Eq.  (\ref{d.densi.2}) contains a
factor $1/(kl)$ and is thus small for wells which are much larger than
a wave length.  Neglecting this term we obtain,

\begin{eqnarray} 
      \bar\rho &=& \frac{2l}{hv} 
      \frac{1+\sin^{2}\tilde{\phi}}{\cos^{2}\tilde{\phi}
      + 4 \tan^{2}\tilde{\phi} \cdot \cos^{2}(\tilde{\theta} +kl)}.   
   \label{d.densi.3} 
\end{eqnarray}
      
\noindent 
As is well known, for the case of opaque barriers 
($\tilde{\theta}\rightarrow \pi/2$ and 
$\tilde{\phi}\rightarrow\pi/2$) the density of 
states becomes a series of delta functions 
   $\lim_{\tilde{r}\rightarrow -1} \bar\rho = \sum_{n=1}^{+\infty} 
   \delta (E - \pi^{2}\hbar^{2} n^{2} / (2ml^{2}))$ 
which coincide with the peaks of perfect transmission. 
For wells much wider than the wavelength, we find from Eq. 
(\ref{d.densi.3}) that the peaks in the density of states 
are at the energies 
   \begin{eqnarray} 
      E_{n} = \frac{\pi^{2}\hbar^{2}}{2ml^{2}} \left\{n 
      -\frac{1}{\pi}\left(\tilde{\theta}-\frac{\pi}{2}\right)
      \right\}^{2} .
   \label{d.peaks} 
   \end{eqnarray}
\noindent 
Here $n$ takes any integer values larger than 
$(2\tilde{\theta}-\pi)/(2\pi)$. 
Comparison of Eq. (\ref{d.derivfried}) with (\ref{d.densi.3})
leads to ${\partial \theta^{\scsc (f)}}/{\partial E} 
      = \pi \, \bar\rho $, i. e. 
the Friedel sum rule in the double-barrier model. 

\begin{figure}[t]
\epsfxsize7.7cm
    \centerline{\epsffile{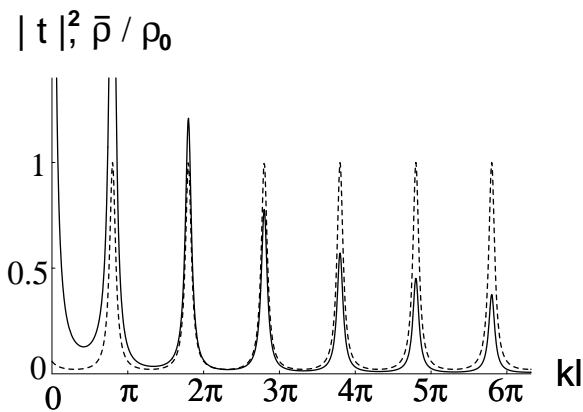}}  
  \caption{Transmission probability (the dashed line) and 
  the density of states $\bar\rho$ (the solid line) in units of  
  $\rho_{0} \equiv ml^{2}/(\pi\hbar^{2})$ as a function of $kl 
  (= l\sqrt{2mE}/\hbar)$ for the double-barrier.} 
  \label{d.transDensi}   
\end{figure}

Using Eqs.
(\ref{d.trans.2}) and (\ref{d.densi.3}) we obtain 

   \begin{eqnarray} 
      \mid t \mid^{2} =  
      \frac{\mid \tilde{t} \mid^{2} }{1 + 
      \mid \tilde{r} \mid^{2}} 
      \frac{\pi \hbar v}{l} \, \bar\rho 
   \label{d.relation} \end{eqnarray}

\noindent 
a relation between the density of state $\bar\rho$ and 
the transmission probability $\mid t \mid^{2}$.
This relation implies that the energy values 
${\cal E}_{n}$ which determine the peaks in the transmission 
probability coincide with the energies $E_{n}$
of the peaks in the density of states. 
Using this fact and  Eq. (\ref{d.densi.3}) we can show 
   
    \begin{eqnarray} 
      \int_{{\cal E}_{n}}^{{\cal E}_{n+1}} dE \; \bar\rho = 1 .
   \label{d.norm} \end{eqnarray}  
\noindent 
In particular, Eq. (\ref{norma}) applies to the 
resonant double-barriers with well widths large 
compared  to the wave length. 
Fig. \ref{d.transDensi} shows the transmission 
probability and the density of state as a function of $kl$ for 
the double-barrier 
for the parameters $\tilde{\theta}=2.2$ and $\tilde{\phi}=2.1$.
We emphasize this behavior of the double barrier, since 
as we show in the next section, a wire with a side branch, exhibits 
a transmission probability and a density of states which do not peak 
at the same energy. 


\subsection{Wire with a side branch} 

\begin{figure}[t] 
\epsfxsize7.7cm
  \vspace{0.5cm}
  \centerline{\epsffile{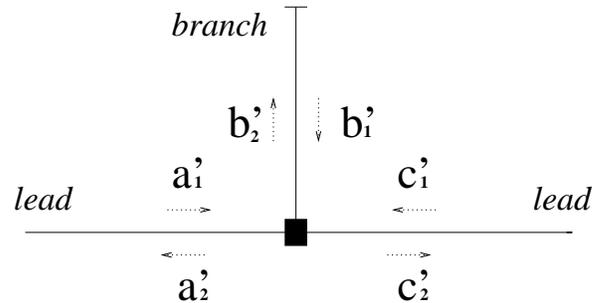}}  
 \vspace{0.7cm}
 \caption{Current amplitudes for the model of a wire with a side branch.}
 \label{stubm} 
\end{figure}

The second example which we investigate is a perfect one-dimensional
wire with a side branch which we also call the stub model.  Such
models have already a long history.  The conductance of a wire with a
side branch was investigated in Refs.  \onlinecite{solo,subra} and
\onlinecite{shao}.  Refs.  \onlinecite{bu93} and \onlinecite{pcmb}
considered charging effects in structures with side branches.  In the
context of the present work the phase behavior of the transmission
amplitude has been investigated by Deo and 
Jayannavar\cite{deo,deo.2}
and Ryu and Cho.\cite{ryu}  However, neither of these two works makes
the distinction between the Friedel phase and the phase of the
transmission amplitude.  Below we investigate these two phases for a
perfect wire to which a side branch of length $l'$ is attached.
A schematic illustration of this system is shown in Fig.  \ref{stubm}.

\noindent 
The junction between the wire and the side branch 
is described by a wave splitter.\cite{ea,buett.2} 
We consider the time-reversal invariant case only
and consider a splitter which is symmetric with respect 
to carriers incident from the left and right lead. 
Furthermore we assume the potential away from the junction 
is the same in all branches. 
The current amplitudes $a'_j$, $b'_j$ and $c'_j$, 
$j=1,2$, which are shown in Fig. \ref{stubm} 
are connected by the following relations,     
     
   \begin{eqnarray} 
      \left( 
      \begin{array}{c} 
         a'_2 \\ b'_2 \\ c'_2
      \end{array}
      \right)
      = 
      \left(
      \begin{array}{ccc} 
         \tilde{r}' & \varepsilon   & \tilde{t}'\\ 
         \varepsilon  & \sigma      & \varepsilon \\ 
         \tilde{t}' & \varepsilon   & \tilde{r}'
      \end{array} 
      \right)
      \left( 
      \begin{array}{c}
         a'_1 \\ b'_1 \\ c'_1
      \end{array}
      \right)
   \label{stub.1}  
   \end{eqnarray}   

\noindent and $b'_1 = \tau' b'_2 \equiv e^{i\varphi'} b'_2 $ with
$\varphi' = 2 k l' + \pi$.  Here $k$ is the wave vector of a particle
with energy $E$.  The constant $\pi$ in $\varphi$ guarantees that the
wave function has a node at the upper end of the stub.  In Eq.
(\ref{stub.1}), $\tilde{r}'$ is the reflection amplitude from a lead
to itself, $\tilde{t}'$ is the transmission amplitude from a lead to
another lead, $\sigma$ is the reflection amplitude from the stub to
itself, $\varepsilon$ is the transmission amplitude from a lead to the
stub or from the stub to a lead, and $\tau'$ is the transmission
amplitude by which the particle starting from the junction returns to
the junction through the stub.

For the scattering matrix, relating the incident amplitudes on the
wire $a'_1 , d'_1$ to the out-going amplitudes $a'_2 , c'_2$ in the
wire (see Fig.  \ref{stubm}) we obtain

\begin{eqnarray} 
      S = \left(
      \begin{array}{cc} 
         \tilde{r}' & \tilde{t}'
         \\ \tilde{t}' & \tilde{r}'
      \end{array} 
      \right)  
      + \frac{\varepsilon^{2}\tau'}{1-\sigma\tau'}
      \left(
      \begin{array}{cc} 
         1 & 1 \\ 1 & 1 
      \end{array} 
      \right). 
    \label{s.smatr} 
\end{eqnarray}

\noindent Here the first matrix on the right hand side arises from
direct transmission past the side branch and direct reflection at the
wave splitter due to the side branch, whereas the second term of the
right-hand side of is the contribution due to carriers which enter the
stub and thus a undergo multiple scattering process.  We assume that
the scattering amplitudes $\tilde{r}'$, $\tilde{t}'$, $\sigma$ and
$\varepsilon$ are real numbers, $\varepsilon\neq 0$ and are
independent of the energy $E$.  These assumptions, and the fact that
the scattering matrix Eq.  (\ref{stub.1}) must be unitary,
demands\cite{buett.2}

\begin{eqnarray}
      && \tilde{r}' = \left( \lambda_{1}
         + \lambda_{2} \sqrt{1-2\varepsilon^{2}} \right)/2 
         \label{s.unita.1} \\
      && \tilde{t}' = \left( -\lambda_{1} 
         + \lambda_{2} \sqrt{1-2\varepsilon^{2}} \right)/2 
         \label{s.unita.2} \\
      && \sigma = -\lambda_{2} \sqrt{1-2\varepsilon^{2}}
         \label{s.unita.3} 
   \end{eqnarray}

\noindent  
where $\lambda_{j}$ is $1$ or $-1$ ($j=1,2$).
Depending on the choice of the $\lambda 's$
four different wave splitters are obtained. 
The value of the coupling constant $\varepsilon$ is in the interval
$[-1/\sqrt{2},1/\sqrt{2} \,]$. Using such a wave splitter leads to a
transmission amplitude

\begin{eqnarray}
      t = \frac{-\lambda_{1} + \lambda_{2} 
      \sqrt{1-2\varepsilon^{2}}}{2}
      \frac{1 + \lambda_{1} e^{2 i kl'}}
      {1 - \lambda_{2} \sqrt{1-2\varepsilon^{2}} \, e^{2 i kl'}}.
   \label{s.trans} 
\end{eqnarray}

\noindent The path of this amplitude in the complex plane as a
function of energy is shown in Fig.  \ref{s.egg} for the case
$\lambda_{1}=-1$.  The path is a circle through the origin since Eq.
(\ref{s.trans}) implies $\mid t + \lambda_{1}/2 \mid^{2} = 1/4$.  From
Eq.  (\ref{s.trans}) it follows that the wire with the stub has zero
points of the transmission probability at the energies

\begin{eqnarray}
      E^{\scsc (n)} = \frac{\pi^{2}\hbar^{2}}{2ml'^{2}} \left(n 
      - \frac{1+\lambda_{1}}{4} \right)^{2},       
   \label{s.transzero} 
\end{eqnarray}

\noindent 
$n = 1, 2,\cdots$. We re-emphasize that 
in contrast to the case of the double barrier,  
the origin is included in the path of the transmission amplitude.



\subsection{Phases and density of states in wire with a side branch} 

Let us now investigate the Friedel phase and the phase of the
transmission amplitude for the wire with a side branch.  Using Eqs.
(\ref{fried}) and (\ref{s.smatr}), the derivative of the Friedel phase
$\theta^{\scsc (f)}$ with respect to the energy $E$ in the wire with a
side branch is given by

\begin{eqnarray} 
      \frac{\partial \theta^{\scsc (f)}}{\partial E} 
      = \frac{\partial\varphi'}{\partial E}
      \left| \frac{\varepsilon}{1-\sigma\tau'} \right|^{2}.  
   \label{s.derivfried} 
\end{eqnarray} 
\noindent 
As shown in the previous subsection there are zero points 
in the transmission amplitude as a function of the energy in 
the stub model, so that abrupt phase changes of the 
transmission amplitude do occur. 
Thus we need a limiting procedure to define 
the phase of the transmission amplitude. 
To this end we add a small perturbation 
$\pm \eta , \eta >0$ to the transmission 
amplitude 
\begin{eqnarray} 
      \bar{t}_{\pm} \equiv t \pm \eta  
   \label{s.transdevia} 
\end{eqnarray} 
and evaluate the phase of ${\bar{t}_{\pm}}$ in the limit  
$\eta \rightarrow +0$.  
The derivative of the phase $\theta^{\scsc (t)}$ 
of the transmission amplitude with respect to the energy $E$ is thus 
given by 
\begin{eqnarray} 
      \frac{\partial \theta^{\scsc (t)}}{\partial E} = 
      \lim_{\eta\rightarrow +0} \left\{1 
      + \left( \frac{\mbox{Im}\{\bar{t}_{\pm}\}}{\mbox{Re}
      \{\bar{t}_{\pm}\}} \right)^{2} \right\}^{-1} 
      \frac{\partial}{\partial E} 
      \frac{\mbox{Im}\{\bar{t}_{\pm}\}}{\mbox{Re}\{\bar{t}_{\pm}\}}.
   \label{s.deriv} 
\end{eqnarray}
\noindent 
Using our specific result for the transmission amplitude 
we obtain 
\begin{eqnarray} 
      \frac{\partial \theta^{\scsc (t)}}{\partial E} 
      =  \frac{\partial \theta^{\scsc (f)}}{\partial E} 
      \pm\lambda_{1} \, \pi \sum_{n=1}^{+\infty} 
      \delta (E - E^{(\scsc n)}). 
   \label{s.deriv.2}
\end{eqnarray}

\noindent 
A detailed derivation of Eq. (\ref{s.deriv.2}) is given in 
Appendix A.
Fig. \ref{s.phase} shows the phase of the transmission 
amplitude of the wire with a side branch 
as a function of $kl'$. We have chosen the branch of the wave splitter
with $\lambda_{1}=-1$, $\lambda_{2}=1$ and a coupling constant 
 $\varepsilon^{2}=0.35$
\begin{figure}
   \epsfxsize7.7cm
\centerline{\epsffile{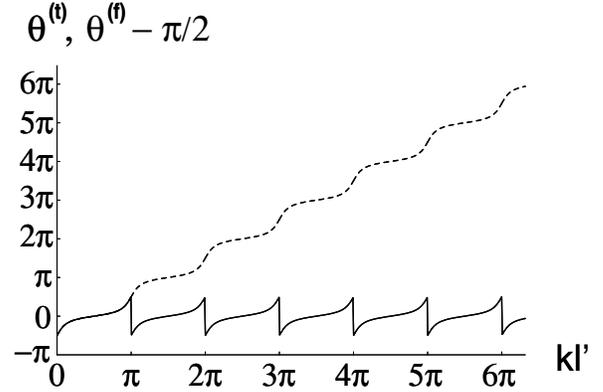}}
\caption{Phase of the transmission amplitude $\theta^{(t)}$
(solid line) and Friedel phase $\theta^{(f)} - \pi/2$ (dashed line)
as a function of $kl'$ for the wire with the side branch.}
\label{s.phase} 
\end{figure}

Let us next investigate the density of states.  The wire is taken to
be on the $x$-axis, with the splitter located at $x= 0$.  The stub
points along the positive $y$-axis.  The splitter is described by an
energy independent scattering matrix and can thus be viewed as point
like.  A scattering state of unit amplitude $\exp\{ik_{\mu}x\}$
describing particles incident from the $\mu$-th lead ($\mu =1, 2$)
gives in the side branch rise to a wave $\Psi_{\mu}(y)$.  Here $k_{1}
= k$ for a wave incident from the left and $k_{2} = -k$ for a wave
incident from the right.  The local density of states $\nu(y)$ in the
side branch is given by $\nu (y) = \sum_{\mu=1}^{2} \mid \Psi_{\mu}(y)
\mid^{2} / (hv)$, and the total density of states is given by the
integral of the local density of states over the entire length of the
stub, $\rho = \int_{0}^{l'} dy \nu(y)$.
We find   

\begin{eqnarray}
      \rho = \frac{\l'}{hv} \left| \frac{2 \varepsilon}{1-\sigma 
      e^{i\varphi'}} \right|^{2} 
      \left( 1 + \frac{\sin\varphi'}{\varphi'
      -\pi} \right). 
   \label{s.densi.2} 
\end{eqnarray}
In the WKB limit of interest here, for a side branch 
much longer than the Fermi wavelength, the second term in  
the bracket of the right-hand side of Eq. (\ref{s.densi.2}) 
can be neglected. 
Using Eq. (\ref{s.unita.3})
we obtain for a long side branch
the density of states
\begin{eqnarray}
      \bar\rho' = \frac{\l'}{hv} 
      \frac{4\varepsilon^{2}}{\left(1-\sqrt{1-2\varepsilon^{2}} 
      \right)^{2} 
      + 4 \sqrt{1-2\varepsilon^{2}} \cdot 
      \sin^{2} (K_{2} l')} .  
   \label{s.densi.3} 
 \end{eqnarray}  
Here, $K_{j}, j = 1,2$ is defined by 
$ K_{j} \equiv  k + (1-\lambda_{j}) \pi / (4l')$. 
For a stub much longer than a wavelength, 
the energies at which the 
density of states peaks are given by 
\begin{eqnarray}
      E'_{n} = \frac{\pi^{2}\hbar^{2}}{2ml'^{2}} 
      \left( n - \frac{1-\lambda_{2}}{4} \right)^{2}. 
   \label{s.densiEnerg} 
\end{eqnarray}
In the week coupling limit $\varepsilon \rightarrow +0$,
using Eq. (\ref{s.densi.3}), we obtain 
$\lim_{\varepsilon \rightarrow +0} \bar\rho' 
      = \sum_{n=1}^{+\infty} \delta (E-E'_{n}).$
Here the right hand-side is the density 
of state of a particle confined in the completely isolated 
stub but taking into account that the phase change 
at the closed coupler is as dictated by 
$\lambda_{2} = 1$. 
The comparison of Eq. (\ref{s.derivfried}) with 
Eq. (\ref{s.densi.3}) leads to
${\partial \theta^{\scsc (f)}}/{\partial E} 
      = \pi \, \bar\rho'$. 
We have thus verified the Friedel sum rule for the wire 
with a side branch.

Let us now show that in this scattering problem 
(for the splitters with $\lambda_{1}\lambda_{2} = -1 $)
the peaks in the 
transmission probability do not co-inside with the peaks 
in the density of states.  
Using Eq. (\ref{s.trans}) the transmission probability $\mid 
t \mid^{2}$ can be expressed in the form, 

   \begin{eqnarray}
      && \mid t \mid^{2} = \nonumber \\
      && \frac{\left(1 - \lambda_{1}\lambda_{2} 
      \sqrt{1-2\varepsilon^{2}}\right)^{2} 
      \cos^{2} (K_{1} l')}
      {\left(1- \lambda_{1}\lambda_{2} 
      \sqrt{1-2\varepsilon^{2}}\right)^{2} 
      + 4 \lambda_{1}\lambda_{2} 
      \sqrt{1-2\varepsilon^{2}} \cdot 
      \sin^{2} (K_{1} l')} .
   \label{s.transproba} \end{eqnarray}
\noindent 
Therefore the energy values ${\cal E}_{n}$, 
$n = 1, 2,\cdots$ at which the transmission probability peaks 
are given by

   \begin{eqnarray}
      {\cal E}_{n} = \frac{\pi^{2}\hbar^{2}}{2ml'^{2}} 
      \left(n - \frac{1-\lambda_{1}}{4}\right)^{2}. 
   \label{s.transenerg} \end{eqnarray}

\noindent 
We see that for  $\lambda_{1}\lambda_{2} = -1$ (i. e. depending 
on the choice of the splitter),  
these energy values are not equal to the energies 
$E'_{n}$, $n = 1, 2,\cdots$ which determine 
the peaks of the density of states. 
Indeed, using Eqs. (\ref{s.densi.3}) and 
(\ref{s.transproba}) we obtain for the relation between density 
of states and the transmission probability
  
   \begin{eqnarray} 
      \mid t \mid^{2} 
      &=& \frac{1}{\lambda_{1} \sigma}
      \left\{  \tilde{t}'^{2} - \frac{\pi \hbar v \varepsilon^{2}}
      {2l'} \bar\rho'  \right\} .
   \label{s.relatDensiTrans} \end{eqnarray}

\begin{figure}[t]    
\epsfxsize7.7cm
\centerline{\epsffile{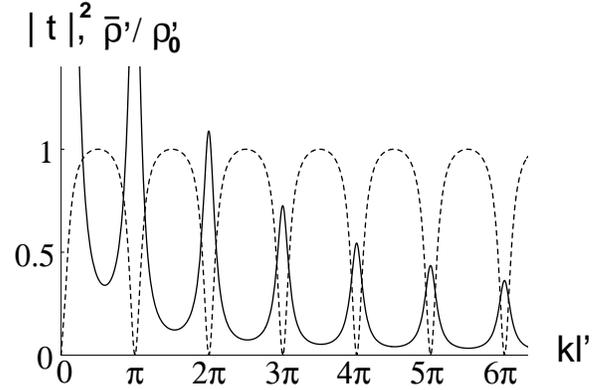}}
\caption{Transmission probability (the dashed line) 
and the density of states $\bar \rho '$ (the solid line) 
in units of $\rho_{0}' \equiv ml'^{2}/(2\pi\hbar^{2})$
for a wire with a side branch.}
\label{s.transDensi} 
\end{figure}

\noindent 
Using Eqs. (\ref{s.densi.3}) and  (\ref{s.transenerg})
we can show that 
$\int_{{\cal E}_{n}}^{{\cal E}_{n+1}} dE \;\; \bar\rho' = 1$. 
This implies that the condition Eq. (\ref{norma}) is fulfilled
also by a wire with a side branch. 
The different behavior, depending on 
the sign of $\lambda_{1}\lambda_{2}$, is especially apparent  
in the week coupling limit. 
In this limit, for  
$\lambda_{1}\lambda_{2} = -1$, almost all the particles incident 
from the wire on the wave splitter 
pass through the junction without noticing the side branch. 
Conversely, in the week coupling limit of a wave splitter with 
$\lambda_{1}\lambda_{2} = 1$ almost all the particles incident 
from the wire on the wave splitter are reflected at the wave 
splitter. 
Fig. \ref{s.transDensi} shows the transmission 
probability and the density of state as a function of $kl'$ for
the wire with a side branch connected by a junction 
with $\lambda_{1} = -1, \lambda_{2} = -1$ and 
$\varepsilon^{2} = 0.35$. 



\subsection{Wire with scattering and a side branch}

\begin{figure}[t] 
\epsfxsize7.7cm
\vspace{0.5cm}
\centerline{\epsffile{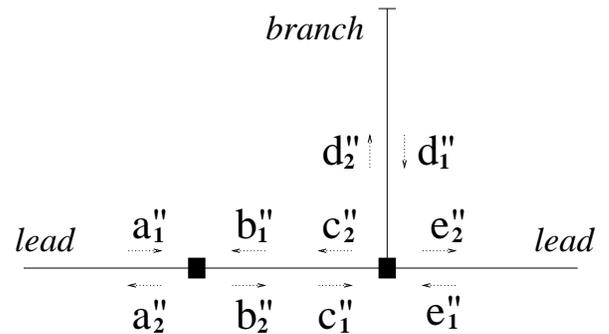}}  
\vspace{0.6cm}
\caption{Current amplitudes for a model with scattering in the wire 
and with a side branch.}
\label{combi} 
\end{figure}

The previous two models are examples which demonstrate 
two different behaviors of the transmission amplitude 
in the complex plane. These different behaviors are illustrated 
by Fig. \ref{peanu} and \ref{egg}. 
Clearly, both of these models are very particular (non-generic)
and the question arises how the behavior exemplified by these
two simple models shows up in more complicated structures. 
To examine this question we now consider a structure 
which incorporates both the resonant double barrier 
and the side branch. 
A schematic illustration of this system is shown in Fig. \ref{combi}.    
\noindent 
We use the same potential barrier as in the resonant double 
barrier structure 
and the same wave splitter as we used for the description
of the wire with the side branch.  Again we will assume that 
the potential outside these scatterers is everywhere the same. 
In this model the current amplitudes $a''_j$, $b''_j$, $c''_j$, 
$d''_j$ and $e''_j$, $j=1,2$ 
which are shown in Fig. \ref{combi} with the 
directions, are connected by the following relations,    
\begin{eqnarray} 
      \left( 
      \begin{array}{c} 
         a''_2 \\ b''_2
      \end{array}
      \right)
      = 
      \left(
      \begin{array}{cc} 
         \tilde{r} & \tilde{t} \\ \tilde{t} & \tilde{r} 
      \end{array} 
      \right)
      \left( 
      \begin{array}{c}
         a''_1 \\ b''_1
      \end{array}
      \right)
   \label{combi.1}  \end{eqnarray} 
   \vspace{-0.5cm}
   \begin{eqnarray}  
   \left( 
      \begin{array}{c} 
         b''_1 \\ c''_1
      \end{array}
      \right)
      = 
      \left(
      \begin{array}{cc} 
         0 & \tau \\ \tau  & 0
      \end{array} 
      \right)
      \left( 
      \begin{array}{c}
        b''_2 \\ c''_2
      \end{array}
      \right)
   \label{combi.2} \end{eqnarray} 
   \vspace{-0.5cm}  
   \begin{eqnarray}
      \left( 
      \begin{array}{c} 
         c''_2 \\ d''_2 \\ e''_2
      \end{array}
      \right)
      = 
      \left(
      \begin{array}{ccc} 
         \tilde{r}' & \varepsilon   & \tilde{t}'\\ 
         \varepsilon  & \sigma      & \varepsilon \\ 
         \tilde{t}' & \varepsilon   & \tilde{r}'
      \end{array} 
      \right)
      \left( 
      \begin{array}{c}
         c''_1 \\ d''_1 \\ e''_1 
      \end{array}
      \right) .
   \label{combi.3}  
\end{eqnarray} 
Furthermore, as in the model with the side branch we put
$d''_1 = \tau' d''_2$.  

For the overall scattering matrix of these system   
we obtain 
\begin{eqnarray} 
      S = \frac{1}{\tilde{r}^{\ast} D} \left(
      \begin{array}{cc} 
         V & W 
         \\ W & -\lambda_{1} \tau'\tau^{2} V^{\ast}  
      \end{array}
      \right) 
    \label{c.smatr} 
\end{eqnarray}
in which $D$, $V$ and $W$ are defined by 
\begin{eqnarray} 
      D \equiv 1 - \sigma \tau' -  \tilde{r} \tilde{r}' 
      (1 + \lambda_{1} \tau') \tau^{2}
   \label{c.smatr.2} \end{eqnarray}    
   \vspace{-0.7cm} 
   \begin{eqnarray} 
      V \equiv D - \mid \tilde{t} \mid^{2} (1 - \sigma \tau') 
   \label{c.smatr.3} \end{eqnarray}    
   \vspace{-0.7cm} 
   \begin{eqnarray} 
      W \equiv \tilde{r}^{\ast} \tilde{t} \, \tilde{t}' 
      (1 - \lambda_{1} \tau') \tau .
   \label{c.smatr.5} 
\end{eqnarray}
The transmission amplitude $t$ can be brought into the form
\begin{eqnarray} 
   t = \frac{\tilde{t} \tilde{t}' (1 + \lambda_{1} 
   e^{2ikl'})e^{ikl}}
   {1 + \sigma e^{2ikl'} - \tilde{r} \tilde{r}' 
   (1 - \lambda_{1} e^{2ikl'}) e^{2ikl} }.
   \label{c.trans} 
\end{eqnarray}
In this representation the transmission amplitude $t$ 
depends on the energy $E$ only through the wave vector 
$k=\sqrt{2mE}/\hbar$. As a function
of energy, the path of this 
transmission amplitude in the complex plane is shown in 
Fig. \ref{c.combin}. 
Here, the parameters chosen for Fig. \ref{c.combin} are 
$\tilde{\theta}=2.2$, $\tilde{\phi}=2.1$, $\lambda_{1}=-1$, 
$\lambda_{2}=1$, $\varepsilon^{2}=0.35$, $l=1$ and $l'=4$.
This figure shows clearly that 
a more general model 
combines the behavior of the 
resonant double barrier model and the stub model. 
Sequences of turns of the transmission amplitude  
which path through the origin are interrupted by 
"double turns" characteristic of the resonant double barrier 
in which the transmission amplitude is non-zero. 
From Eq. (\ref{c.trans}) it can be noted that the wire with 
scattering and with a side branch 
has the same zero points for the transmission
amplitude as the wire with the side branch.


\section{Conclusions} 

In this work we have discussed the transmission amplitude as 
function of energy in the complex plane for scattering systems 
(without a magnetic field) connected 
to two single channel leads. We emphasize that there are two phases
which are of importance, namely the phase which appears in the 
Friedel sum rule and the phase of the transmission amplitudes. 
Except in special cases, these two phases are in general different. 
This important point has also been emphasized by Lee.\cite{Lee} 
The two phases are different
if the transmission amplitude exhibits an energy 
at which it is zero. At these energies the transmission amplitude 
phase and the Friedel phase acquire an additional difference 
given by  $\pm\pi$. If the transmission amplitude exhibits 
no zero point between resonances the variation of the phase 
from one resonant peak to another is close to $\pi$.
If a zero point exists the phase change between   
consecutive resonance peaks of the transmission probability is 
close to $2\pi$.  
This difference is shown clearly in the paths 
of the transmission amplitude in the complex plane. 

For a sufficiently general model, we expect sequences of 
resonant peaks which are in phase (the phase increases by $2\pi$)
interrupted by peaks which are out of phase (the phase increases
only by $\pi$ as we go from one peak to the next). In terms 
of the parity of the Aharonov-Bohm effect, these implies sequences
of peaks over which the parity is conserved. These sequences
are interrupted 
by transitions which generate a flip in the parity. 
It is clearly desirable to investigate now a number of statistical 
questions: For example for a fully chaotic quantum dot 
one would like to find the ensemble averaged density of zero's 
and compare this with the ensemble averaged density of states. 
Furthermore, one would like to know if such a cavity 
exhibits correlations in the occurrence of zero's
(long sequences 
of zero's interrupted by a flip in the parity), etc. 
Since the distribution of eigenvalues is known, 
random matrix theory likely gives an answer to these questions. 

We add a remark on Fano resonances: Fano resonances arise due to the 
coupling of discrete states with continuous states.  Such 
resonances also exhibit transmission 
zero's.\cite{fano,noek,goeres} Consequently, 
for such resonances the Friedel phase also does not 
coincide with the phase changes of the transmission amplitudes. 
The wire with a side branch investigated here also couples 
a set of discrete states with a continuum and thus provides 
for interfering transmission paths. However, the resonances in this 
case are not of the Fano type (as shown in Fig. \ref{s.transDensi})
but rather Breit-Wigner resonances in the reflection probability. 

It is very interesting to investigate the behavior of the 
transmission amplitude in the complex plane for variations in 
parameters other than the energy. For instance we can ask about 
the path of the transmission amplitude if the AB flux increases 
by a flux quantum in a multiply connected geometry. The additional 
questions raised in this section clearly demonstrate that the 
investigation of the path of the transmission amplitude in the 
complex plane is an interesting avenue of future research. 


\acknowledgments
We are grateful to A. Levy Yeyati, M. Devoret and D. Esteve 
for stimulating discussions. 
T.T acknowledges the support of the Swiss Federal government.
M. B. is supported by the Swiss National Science foundation. 

\appendix 
\section{Relation between phases 
for the wire with a side branch}

%
In this Appendix we outline the derivation of Eq. (\ref{s.deriv.2})
which relates the Friedel phase and transmission 
amplitude phase for the wire with a side branch. 
First, using Eqs. (\ref{s.trans}) and 
(\ref{s.transdevia}) we find
\begin{eqnarray}
      \bar{t}_{\pm} = \frac{-\lambda_{1} - \sigma}{2}
      \frac{1 - \lambda_{1} e^{i \varphi'}}{1 - \sigma \, 
      e^{i \varphi'} } \pm\eta . 
   \label{a.trans} 
\end{eqnarray}

The transmission amplitude $t$ is dependent on the 
energy only through the quantity $\varphi'$, so that we obtain 
\begin{eqnarray} 
      \frac{\partial \theta^{\scsc (t)}}{\partial E} = 
      \lim_{\eta\rightarrow +0} 
      \frac{\partial \varphi'}{\partial E}
      \frac{\partial \mbox{Arg}\{\bar{t}_{\pm}\}}{\partial 
      \varphi'}.  
   \label{a.deriv.1} 
\end{eqnarray}

\noindent 
Using Eq. (\ref{a.trans}) the derivative 
$\partial \mbox{Arg}\{\bar{t}_{\pm}\}/\partial \varphi'$ is 
found to be 
\begin{eqnarray} 
      && {\displaystyle \frac{\partial 
      \mbox{Arg}\{\bar{t}_{\pm}\}}{\partial \varphi'}} 
      \nonumber  \\
      && = \left\{1 + \left( \frac{\mbox{Im}\{\bar{t}_{\pm}\}}
         {\mbox{Re}\{\bar{t}_{\pm}\}} \right)^{2} \right\}^{-1} 
         \frac{\partial}{\partial \varphi'} 
         \frac{\mbox{Im}\{\bar{t}_{\pm}\}}
         {\mbox{Re}\{\bar{t}_{\pm}\}}
         \nonumber \\
      && = \frac{1}{2} 
         \frac{1-\sigma^{2}}{1+\sigma^{2}
            - 2\sigma\cos\varphi'} 
      \nonumber \\
      && \hspace{0.5cm} \times
         \frac{\mp 2 f_{1}(\sigma,\varphi') 
            \eta+f_{3}(\sigma,\lambda_{1},\varphi')}
         {2f_{2}(\sigma,\varphi') \eta^{2}
         + (1\mp 2\lambda_{1}\eta) 
         f_{3}(\sigma,\lambda_{1},\varphi')} 
         \nonumber \\
      && = \frac{1}{2} 
         \frac{1-\sigma^{2}}{1+\sigma^{2}
            - 2\sigma\cos\varphi'} \nonumber \\
      && \hspace{0.5cm} + \frac{1}{2} 
         F\left( 1 \mp 2\lambda_{1}\eta, \,\lambda_{1}\sigma, \, 
         \varphi'+\frac{1+\lambda_{1}}{2}\pi \right) 
   \label{a.deriv.2} 
\end{eqnarray}

\noindent 
where the functions $f_{1}(x,y), f_{2}(x,y)$ 
and  $f_{3}(x,y,z)$ are defined by 
\begin{eqnarray} 
      f_{1}(x,y) \equiv 2x-(1+x^{2})\cos y
   \label{a.a1} \end{eqnarray}
   \vspace{-0.7cm} 
   \begin{eqnarray} 
      f_{2}(x,y) \equiv 1+x^{2}-2x\cos y
   \label{a.a2} \end{eqnarray}
   \vspace{-0.7cm} 
   \begin{eqnarray} 
      f_{3}(x,y,z) \equiv (1+ xy)^{2}(1
            - y \cos z)
   \label{a.a3} 
\end{eqnarray}
   
\noindent 
and the function $F(x,y,z)$ is 
defined by 
\begin{eqnarray} 
      && F(x,y,z) \nonumber \\  
      && \equiv \frac{(1-x^{2})(1-y^{2})}
      {(1-x)^{2}(1-y)^{2}+2(1+xy)(x+y)(1+\cos z)}.
   \label{a.funct.F1} \end{eqnarray}
   
\noindent 
$F(x,y,z)$ has the following properties. 
First it follows that 
\begin{eqnarray} 
      && \lim_{x \rightarrow 1 \pm 0}  
      F(x,y,z) \nonumber \\  
      && \hspace{0.2cm} = \left\{ 
      \begin{array}{cl}
      0 & \mbox{in} \; z=(2n+1)\pi \\
      \mp \epsilon (1-y^2) \times \infty 
      & \mbox{in} \; z \neq (2n+1)\pi, 
      \end{array}
      \right.
   \label{a.funct.F2} 
\end{eqnarray}
$n=0,\pm 1,\pm 2,\cdots$. 
Second it follows that
\begin{eqnarray} 
      \hspace{0cm}&& \int_{\lambda}^{\lambda+2\pi} dz 
      \, F(x,y,z) \nonumber \\ 
      &&= \frac{1}{i} \oint_{C} d\omega \frac{(1-x^{2})(1-y^{2})}{
      \left\{(1+xy)\omega+x+y \right\} \left\{
      (x+y)\omega+1+xy \right\}} \nonumber \\ \nonumber \\
      &&= \left\{ 
      \begin{array}{cl}
      + 2\pi & \mbox{in} \;\; 0 \neq \, 
      \mid x+y \mid < \mid 1+xy \mid  \\
      - 2 \pi & \mbox{in} \; \mid x+y 
      \mid > \mid 1+xy \mid \, \neq 0
      \end{array}
      \right.
   \label{a.funct.F3} 
\end{eqnarray}
   
\noindent where $\lambda$ is a real number
and $\omega \equiv \exp (iz)$.  
$C$ is the path running 
counterclockwise on the circle whose center is the origin 
and the radius is 1.
Eqs. (\ref{a.funct.F2}) and (\ref{a.funct.F3}) lead to 
\begin{eqnarray} 
      && \lim_{x \rightarrow 1\pm 0} F(x,y,z) \nonumber \\ 
      && \hspace{0.2cm} = \mp 2\pi \epsilon (1-y^2) 
      {\displaystyle \sum_{n=-\infty}^{+\infty}} 
      \delta (z-(2n+1)\pi). 
   \label{a.delta} 
\end{eqnarray}
Using Eq. (\ref{s.transzero}) and the inequality $E>0$, 
we obtain  
\begin{eqnarray} 
      \delta \Biggl(\varphi' +\frac{1+\lambda_{1}}{2}\pi 
      && -(2n+1)\pi \Biggl) 
      = \left(\frac{\partial\varphi'}{\partial E} 
      \right)^{-1} \delta(E-E^{(\scsc n)}).  \nonumber \\
      &&
   \label{a.delta.2} 
\end{eqnarray}
From Eqs. (\ref{a.deriv.1}), (\ref{a.deriv.2}), (\ref{a.delta}) and 
(\ref{a.delta.2}) we derive 

  \begin{eqnarray} 
      \frac{\partial \theta^{\scsc (t)}}{\partial E} 
      =  \frac{\partial\varphi'}{\partial E}
      \left| \frac{\varepsilon}{1-\sigma\tau'} \right|^{2}  
      \pm\lambda_{1} \, \pi \sum_{n=1}^{+\infty} 
      \delta (E - E^{(\scsc n)}). 
   \label{a.derivative} \end{eqnarray}

\noindent Using Eqs. (\ref{s.derivfried}) and (\ref{a.derivative}) 
we obtain Eq. (\ref{s.deriv.2}).




\end{multicols}

\end{document}